\documentclass[a4paper,11pt]{article}

\pdfoutput=1 

\usepackage{jheppub}

\usepackage{mathrsfs}
\usepackage{comment}
\usepackage{centernot}
\usepackage{mathtools}
\usepackage{stmaryrd}

\usepackage{latexsym,amssymb,amstext,amsmath}
\usepackage{hyperref}
\usepackage{graphbox,graphicx}
\usepackage{amsthm}
\usepackage{esint}
\usepackage{floatrow}

\usepackage{footmisc}

\def\bea {\begin{eqnarray}}
\def\eea {\end{eqnarray}}

\def\beq{\begin{equation}}
\def\eeq{\end{equation}}
\def\beqa{\begin{eqnarray}}
\def\eeqa{\end{eqnarray}}

\theoremstyle{definition}

\title{\boldmath Exact BPS black hole microstate counting from holographic conformal quantum mechanics}

\author{Gabriel Lopes Cardoso${}^1$ and Suresh Nampuri${}^2$}

\affiliation{${}^1$ Center for Mathematical Analysis, Geometry and Dynamical Systems,\\
Department of Mathematics, Instituto Superior T\'ecnico, Universidade de Lisboa,\\
Av. Rovisco Pais, 
1049-001 Lisboa, Portugal}
\affiliation{${}^2$ Center for Mathematical Studies,	
Faculdade de Ci\^encias, Universidade de Lisboa,\\
Edf. C6, 
Campo Grande, 1749-016 Lisboa, Portugal }

\emailAdd{gabriel.lopes.cardoso@tecnico.ulisboa.pt}
\emailAdd{nampuri@gmail.com}

\abstract{In this note, we develop a prescription for describing BPS black hole microstates in terms of a holographic conformal quantum mechanics (CQM) model dual to the near-horizon $AdS_2$ geometry of the black hole. We use 1/2 BPS small black holes in a 4D ${\cal N}=4$ toroidal heterotic compactification as well as 1/8 BPS large black holes in a 4D ${\cal N}=8$ Type II toroidal compactification as test cases for our approach. In each case, the $SL(2,\mathbb{Z})$ modular symmetries of the known generating function of the exact microstate degeneracies enables the latter to be expressed as a Rademacher series expansion, with each summand consisting of phases and a modified Bessel function of the first kind. We make a motivated ansatz that the de Alfaro-Fubini-Furlan model (DFF) coupled to a simple harmonic oscillator is a universal sector of the holographic CQM dual to the BPS black hole's near-horizon $AdS_2$ geometry, and demonstrate how in both cases, the two parameters of this putative CQM, the DFF coupling as well as the oscillator frequency, exactly encode both the index and the argument of the Bessel function. Consequently, we extract the leading, logarithmic and all sub-leading power law black hole entropy contributions from calculations in the CQM. In the ${\cal N}=4$ case, the  DFF ansatz is sufficient to successfully reproduce the exact microscopic generating function from the CQM.

 }

\begin{document} 
\maketitle
\flushbottom

\section{Introduction}
The exact counting of the statistical microstates for certain classes of BPS black holes, modeled microscopically in terms of a bound state of strings and branes wound on compact manifolds \cite{Sen:1995in,Strominger:1996sh}, has been one of the keystone computations of superstring theory along its journey to establish its viability as a consistent theory of quantum gravity. Concurrently, the AdS/CFT holographic correspondence \cite{Maldacena:1997re} encodes one of the deepest and most incisively insightful blueprints for offering a dual description of quantum gravity in  AdS in terms of a holographically dual conformal field theory. Under this correspondence, black hole configurations in AdS, in particular, are viewed as thermal descriptions in the holographically dual CFT. This approach has proved to be particularly useful in understanding non-perturbative dynamics such as phase transitions in black holes \cite{Witten:1998zw} of a more general class than BPS,  and for black hole counting, specifically, the computation of the Bekenstein Hawking entropy for such black holes \cite{Strominger:1997eq,Maldacena:1998bw}. However, so far, there have been no exact counting results under the aegis of this correspondence. This note takes steps towards bridging the stringy microstate and holographic modeling approaches to black holes by systematically interpreting exact counting results from the former in terms of computations or quantities in the latter.

We will focus on the 4D $\tfrac12$ BPS small black holes in a ${\cal N}=4$ toroidal heterotic compactification \cite{Sen:1995in} as well as 4D $\frac18$ BPS large black holes arising in a ${\cal N}=8$ toroidally compactified Type II superstring theory \cite{Maldacena:1999bp}, with vanishing and non-zero classical horizon areas respectively.
Both kinds of black holes incorporate an $AdS_2$ factor in their near-horizon geometry and consequently are, under the ambit of the AdS/CFT correspondence, amenable to being represented as states in a holographically dual CFT or Conformal Quantum Mechanics (CQM). Further, both the $\tfrac12$ BPS and $\tfrac18$ BPS black holes have exact degeneracy formulae based on a microstate string picture \cite{Dabholkar:1995nc} and given by $1/\eta^{24}(\tau)$ \cite{Dabholkar:2005dt}, 
the inverse of the $24$th power of the Dedekind eta function, a weakly holomorphic modular form of weight $-12$, and the weight $-2$ Jacobi form $\vartheta_1^2(\tau,z)/\eta^6(\tau)$ \cite{Maldacena:1999bp}, both defined on the upper half plane parametrised by the complex variable $\tau$ and which transform covariantly under the $SL(2,\mathbb{Z})$ group. The integral degeneracies are then the Fourier or Fourier-Jacobi coefficients of the generating function expanded around $\tau = i \infty$. The modular symmetries imply that given a finite number of Fourier coefficients (polar coefficients), any other Fourier coefficient and therefore the generating function itself can be written as a convergent infinite sum over the countably infinite set of rationals in $[0,1)$, with each summand constituted of a modified Bessel function of the first kind,  polar term(s) and phase factors. This summation is referred to as the Rademacher series \cite{rademacher1937convergent} expansion of the Fourier coefficient \cite{hardy1918asymptotic} and it encodes a core insight about the organisation of quantum microstates in the gravitational configuration under consideration --  namely that all the data or information in a non-perturbative configuration such as a BPS black hole in this theory of quantum gravity can be rigorously determined by specifying a finite amount of data (polar terms), 
which encode the fundamental degrees of freedom in the theory.  The modified Bessel functions appearing in the Rademacher series expansion have an index related to the weight of the corresponding modular or Jacobi form generating function. Hence modular symmetries are powerful enough to reduce the problem of encoding a count of BPS microscopic degeneracies in terms of two sets of numbers -- polar coefficients and the index of the modified Bessel function. 

For encoding BPS entropy from a AdS/CFT(CQM) perspective, we will approximately model the BPS black hole with 
near-horizon geometry $AdS_2 \times S^2$ as a multi-species collection of $N\gg 1$ 
 BPS  particles moving in the near-horizon Poincar\'e patch $AdS_2$ background, with additional quantum numbers/degrees of freedom associated with the additional $S^2$. In \cite{Claus:1998ts} as well as later in \cite{Gibbons:1998fa}, it was shown that the resultant worldline Hamiltonian was that of a Calogero model. In order to obtain a discrete spectrum upon quantisation, we follow \cite{Gibbons:1998fa} and 
  we shift from Poincar\'e to compact black hole/Rindler time (cf. \eqref{rindads}), thereby introducing a simple harmonic oscillator coupling 
 in this model.  
 The ground state wave functions peak at the boundary of the Poincar\'e patch $AdS_2$ background \cite{Lechtenfeld:2015wka}, and hence the particles are all stuck near the boundary, indicating a sector of the boundary dynamics governed by a 
de Alfaro-Furlan-Fubini (DFF) type model \cite{deAlfaro:1976vlx}. (For more on how the DFF model and in general conformal actions emerge in the context of near-horizon $AdS_2$, cf.  \cite{Gibbons:1998fa, Michelson:1999dx, Cacciatori:1999rp, Cadoni:2000gm,Verlinde:2004gt}).  
This suggests that the holographic boundary dynamics encoding bulk physics will include a collective contribution from $s$ species of aggregate effective particles, each composed of a large number $N$ of particles localised near the boundary as well as the aggregate particles' quantum mechanics. We will see that this is indeed borne out in our holographic model proposal wherein the dominant Bekenstein-Hawking exponential degeneracy is associated with a collective degree of freedom, while the subleading corrections are  encoded in a Euclidean heat kernel trace in the conformal quantum mechanics of a single particle. 

This motivates us to postulate a universal sector of the boundary CQM as being 
a model of DFF particles belonging to $s$ different, mutually non-interacting species, each consisting of a very large number $N$ of DFF particles
coupled to a simple harmonic oscillator. For each species, the associated Hamiltonian is given by
   \bea {\mathcal H} = \sum_{i=1}^N \left( \frac{p_i^2}{2 m} + \frac{g}{x_i^2} + \frac{1}{2} m \omega^2 x_i^2 \right)\;\;\;,\;\;\;
   g> 0 \;.
\label{Ndff}
\eea
The above Hamiltonian can be seen as representing a set of $N$ CQM models with identical couplings, each of which corresponds to a DFF particle coupled to a simple  harmonic oscillator. The static solution for each CQM is therefore the same value $x_0$, as given below in \eqref{x04}.
Note that $\mathcal H$ only depends on two couplings, $g$ and $\omega$.
In the $N \rightarrow + \infty$ limit, the generating function of degeneracies of the spectrum becomes a generator of integral partitions and 
is given by 
$1/\eta^s(\tau)$, a weakly holomorphic form of weight $-s/2$  \cite{Lechtenfeld:2015wka}. For the case of $\tfrac12$ BPS ${\cal N}=4$ small black holes, the choice of $s=24$ yields the exact microstate generating formula, while for the case of $\tfrac18$ BPS ${\cal N}=8$ large black holes, we obtain an approximation to the Jacobi form generating the microstate degeneracies for the choice of $s=5$, in that the Rademacher series expansion of the approximate formula exactly reproduces the leading, logarithmic and sub-leading terms in a perturbative large charge expansion of the black hole entropy, though not the Kloosterman phases in the Rademacher series expansion of the exact microstate degeneracies.

Allowing for the presence of $s$ species of DFF particles is crucial for reproducing  the leading large charge BPS entropy as well as the subleading logarithmic correction to it.  We also  show that the trace of the Euclidean heat kernel in the DFF model, augmented by the harmonic oscillator term in \eqref{Ndff}, encodes all the sub-leading power law corrections to the BPS entropy in a perturbative large charge expansion in both cases.
The DFF parameters, $g$ and $\omega$, are linked to the index $a$ of the modified Bessel function of the first kind appearing in the Rademacher series expansion, and therefore to the weight 
\bea
\frac{s}{2}=a-1
\label{as}
\eea
of the corresponding generating function,  $1/\eta^s(\tau)$, as well as the argument $z$ of the Bessel function via 
\bea
 \frac{2 m g}{\hbar^2}  = 
a^2 - \frac14   \;\;\;,\;\;\;
z = \frac{\sqrt{a^2 - \frac14}}{\gamma\omega } \;.
\eea 
In the above, $\gamma$ runs over the natural numbers and, from the microscopic degeneracy perspective, labels distinct sectors in the Rademacher expansion, while in the CQM picture, it is identified with the periodicity of Euclidean time in the DFF action (cf. \eqref{Zbe}), corresponding to each such sector. 
The sub-leading power law corrections to the BPS entropy mentioned above are captured by ${\tilde K_E}(z, a)$, the Euclidean heat kernel  in the limit $\omega  \rightarrow 0$, 
\bea
{\tilde K_E}(z, a) 
= \frac{1}{\sqrt{2 \pi}} \left( 
 1 - \frac{(4 a^2 -1)}{8 z} + 
\frac{(4 a^2 -1)(4 a^2 -3^2)}{2! (8z)^2} - \dots \right)  \;.
 \eea

Hence, the core result of this paper is to establish that in the $N\rightarrow + \infty$ limit, a set of $N$ CQM models, as  given in  \eqref{Ndff}, for each of the $s$ species  is a reliable approximation to the holographic CQM dual to the near-horizon $AdS_2$ geometry corresponding to 4D $\frac{1}{2}$ BPS and $\frac{1}{8}$ BPS black holes in $\mathcal{N}=4$ and $\mathcal{N}=8$ string theories, respectively. Similar attempts involving the Farey tail expansion of the microscopic degeneracy formulae have been made in the $AdS_3/CFT_2$ contexts in \cite{Dijkgraaf:2000fq,Manschot:2007ha}.

This note is structured as follows. In Section \ref{CQM} we begin
with a review of the relevant features of the DFF model \cite{deAlfaro:1976vlx}, as well as its quantisation, after coupling to a harmonic oscillator term of frequency $\omega$. We will refer to the latter as DFF$_\omega$. This serves as a useful backdrop for what follows, namely demonstrating how microscopic counting formulae for $\tfrac12$ BPS and $\tfrac18$ BPS black holes in ${\cal N}=4$ and ${\cal N}=8$ compactifications respectively arise, exactly or approximately, from a holographic DFF$_\omega$ CQM perspective. This is done in Section \ref{Embedding}. We conclude with a summary of results and comments on future directions. In Appendix \ref{sec:heat}
we review features of the path integral representation of the heat kernel in quantum mechanics, and in Appendix \ref{sec:lim}
we review various limiting cases of the 
Euclidean heat kernel for the model described by the Euclidean Lagrangian \eqref{euclLag}.

\section{Conformal quantum mechanics: the DFF and DFF$_\omega$ models}\label{CQM}

In this section we summarise various features of the DFF and DFF$_\omega$ model.

\subsection{The DFF model}

The DFF model, a conformal quantum mechanics model  \cite{deAlfaro:1976vlx}, is described by the 
Hamiltonian
\bea
H = \frac12 \left( p^2 + \frac{{\tilde g}}{x^2} \right) \;\;\;,\;\;\; {\tilde g} > 0 \;,
\label{HDFF}
\eea
where ${\tilde g}$ denotes the dimensionless coupling of the model. Here we follow the detailed studies 
\cite{Chamon:2011xk,Jackiw:2012ur}.

The underlying  $so(2,1)$ algebra can be canonically realized in terms of operators $H, \mathcal{K}, D$, where $H$ denotes the Hamiltonian, $\mathcal{K}$ the conformal boost generator and $D$ the dilation generator. It can also be realized in terms of the generators  $R, L_{\pm}$ given by
\bea
R = \frac12 \left( \frac{\mathcal{K}}{A} + A \, H \right) \;\;\;,\;\;\; L_{\pm } = \frac12 \left( \frac{\mathcal{K}}{A} - A \, H \right) \pm i \, D \;,
\eea
where $A$ denotes a constant with the appropriate scaling dimension. These generators satisfy
\begin{equation}
    [R, L_\pm] = \pm  L_\pm\,, \qquad [L_-, L_+] = 2 R \,.
    \label{daffalg}
\end{equation}
While $H$ has a continuous spectrum with $E >0$ (but no ground state at $E=0$), $R$ 
has a discrete spectrum with a unique ground state $|0\rangle$,
\bea
R |n \rangle = \left( n + r_0 \right) \, |n \rangle \;\;\;,\;\;\; n \in \mathbb{N}_0 \;\;\;,\;\;\; \langle  m | n \rangle  = \delta_{m,n} \;,
\label{Rgen}
\eea
where 
$r_0$ is expressed in terms of the coupling constant ${\tilde g}$ of the model as
\bea
r_0 = \frac12 (1 + {\tilde a} ) \;\;\;,\;\;\; {\tilde a} =  \frac12   \left( 1 + \frac{4 {\tilde g}}{\hbar^2} \right)^{1/2}  \;,
\eea
where we have reintroduced the dependence on $\hbar$. Setting \cite{Akhoury:1984pt}
\bea
{\tilde g} = {\tilde \alpha} ({\tilde \alpha} + 1) 
\hbar^2
\label{tga}
\eea
(with ${\tilde \alpha} >0$ to ensure ${\tilde g} >0$), we obtain
\bea
{\tilde a} = {\tilde \alpha} + \frac12 \;.
\label{aal}
\eea

At $t=0$, $\mathcal{K}$ can be represented by $\frac12 x^2$, in which case 
\bea
R =  \frac{A}{4}  \left( p^2 + \frac{{\tilde g}}{x^2} \right) + \frac{1}{4A} x^2 \;.
\eea
Setting $A = 1/(m \omega)$, we obtain 
\bea
R = \frac{1}{2 \omega}  \left(  \frac{p^2}{2m}  +  \frac{g}{x^2} + \frac12 m \omega^2 x^2 \right) \;,
\label{genR}
\eea
where
\bea
g = \frac{{\tilde g}}{2m} \;,
\eea
and hence
\bea
r_0 =  \frac12 (1 + {\tilde a} ) \;\;\;,\;\;\; {\tilde a} =    \frac12 \left( 1 + 
\frac{8 m g}{\hbar^2} \right)^{1/2}  \;.
\label{r0arel}
\eea

Regarding $ 2 \omega R$ as a Hamiltonian (cf. \eqref{Ndff}), we associate to it the following Euclidean DFF$_\omega$ Lagrangian,
\bea
L_E(x, \dot{x}) = \frac12 m \dot{x}^2 + V(x) \;\;\;,\;\;\; V(x) = \frac{g}{x^2} + \frac12 m \omega^2 x^2  \;\;\;,\;\;\; g > 0 \;,
\label{euclLag}
\eea
where we restrict to the half line $x>0$.
We assign the following scaling dimensions (which we denote by  $[\cdot ]$)
to the various quantities in the Lagrangian,
\bea
[t] = 1 \;\;\;,\;\;\; [x] = \frac12 \;\;\;,\;\;\; [m] = 0  \;\;\;,\;\;\; [\omega] = -1 \;\;\;,\;\;\; [g] = 0 
\;\;\;,\;\;\; [\hbar] = 0 \;.
\label{scal}
\eea
Then $A = 1/(m \omega)$ has scaling dimension $1$, 
$L_E$ has scaling dimension $-1$ and the action $S_E = \int dt L_E $ is invariant under scalings. Note that $R$ and $r_0$ have scaling dimension zero.

\subsection{The heat kernel }

Now we discuss the heat kernel $K_E$ for the system described by the Euclidean Lagrangian \eqref{euclLag}. 
We refer to Appendix \ref{sec:heat} for a summary of the path integral representation of the heat kernel.

Remarkably,
the heat kernel $K$ for this system\footnote{The heat kernel $K$ should not be confused with the conformal boost generator $\mathcal{K}$.} was exactly computed in Minkowski time long time ago in \cite{Khandekar} using a path integral representation. Recent detailed studies via path integrals can be found in \cite{Chakraborty:2022qdr,Camblong:2022oet,Camblong:2024bsy,Camblong:2024jpq}. An early application of the heat kernel $K$ in the context of Rindler space string thermodynamics appeared in \cite{Mertens:2013zya}.
In \cite{Mertens:2013zya} the coupling $g$ was taken to be negative, whereas here $g>0$.
By Wick rotating to Euclidean time and
performing the following replacements in their expression for the heat kernel,
\bea
i \sin \phi \rightarrow \sinh \phi \;\;\;,\;\;\; \cos \phi \rightarrow \cosh \phi \;\;\;,\;\;\;
\dot{s} = 0 \;\;\;,\;\;\; \dot{\gamma} = \omega \;\;\;,\;\;\;
\phi = \omega (t_f - t_i) \;,
\eea
we obtain for the Euclidean heat kernel, 
\bea
K_E (y, t_f; x, t_i) = \frac{m \omega }{\hbar}\frac{\sqrt{ x y }}{ \sinh  \omega (t_f - t_i) }  \; e^{- \frac{m \omega}{2 \hbar} ( x^2 + y^2)  \coth  \omega (t_f - t_i)} \, I_a \left( \frac{m \omega }{\hbar} 
\frac{ x y }{\sinh  \omega (t_f - t_i)} \right) \;, \nonumber\\
\label{heatom}
\eea
where $I_a (z)$ denotes the modified Bessel function of first kind. 
The dependence on the coupling $g$ is contained in the index $a$ of the modified Bessel function \cite{Khandekar} and reads
\bea
 a =    \frac12 \left( 1 + \frac{8 m g}{\hbar^2} \right)^{1/2} \;.
  \label{indamg}
 \eea

Now consider the scaling dimensions given in \eqref{scal}. 
We see that $K_E$ given in \eqref{heatom}
has scale dimension $-1/2$: the arguments of the Bessel function and of the exponent have scaling weight $0$, while the prefactor has scaling weight $-1/2$. The integrated heat kernel given below in \eqref{heatpath2} has scaling weight zero.

Next, let us focus on periodic paths $\gamma$ that start and end at the same point $x$, i.e. $x_i(t_i) = x_f (t_i + T) = x$ with $T = t_f - t_i$.
Then, the associated heat kernel, which we denote by $K_E (x, x, T)$, takes the form
\bea
K_E (x, x, T) =  \frac{m \omega }{\hbar} \frac{x }{ \sinh  \omega T }  \; e^{- \frac{m \omega}{ \hbar}  x^2  \coth  \omega T} \, I_a \left( \frac{m \omega }{\hbar} 
\frac{ x^2 }{\sinh  \omega T } \right)
\;.
\label{heatomx0}
\eea
We will set $T$ in the above equation to a value consistent with the physics of the system. We will see this explicitly in Section \ref{Embedding} wherein we identify $T$ with the index $\gamma$ labelling distinct sectors in the microscopic degeneracy formulae of certain BPS black holes, when relating the couplings of the DFF$_\omega$ model to the data appearing in these formulae.

\subsection{The partition function \label{sec:pf}}

The partition function for the Hamilton operator  $ 2 \omega R$ can be computed in a straight forward manner using \eqref{Rgen},
\bea
Z(T) = {\rm Tr} e^{- 2 \omega T R} = 
\sum_{n \in \mathbb{N}_0} 
 e^{-(n + r_0) 2 \omega T } = \frac{e^{(1-r_0) 2 \omega T }}{e^{2 \omega T } -1} \;.
  \label{trR}
\eea
The partition function can also be obtained using the heat kernel $K_E$, as shown in \cite{Camblong:2024jpq}.
Namely, {from} \eqref{heatZ} we infer
\bea
Z(T) &=& \int_0^{+\infty}  K_E(x,x,T)  \, dx\;,
\label{heatpath2}
\eea
where 
we recall that we have to restrict to $x> 0$. Using \eqref{heatomx0}, 
changing the integration variable from $x$ to $v =  
\frac{m \omega}{ \hbar}  x^2 $
and using eq. (B.17) in \cite{Camblong:2024jpq},  we obtain
\bea
\label{zpath}
 \int_0^{+\infty}  K_E(x,x,T) \, dx &=& \frac{1}{2 \sinh \omega T} 
 \int_0^{+\infty} e^{- v  \coth \omega T} \, I_a \left( 
\frac{ v }{\sinh \omega T } \right) dv  \nonumber\\
&=& \frac{1}{2 \sinh \omega T}  \, e^{- a \omega T} = \frac{e^{(1-a) \omega T}}{e^{2 \omega T} - 1} \;.
\eea
By comparing \eqref{zpath} with \eqref{trR}, we infer the relation
\bea
a = 2 r_0 -1 \;,
\label{ar0}
\eea
which relates the index $a$ of the modified Bessel function $I_a$ to the quantity $r_0$ given in \eqref{r0arel}. {From} the latter we infer that ${\tilde a} = a $, which is consistent and agrees with \eqref{indamg}.

As an application, consider the case when $r_0=1$, which corresponds to $a=1$ and to ${\tilde \alpha} = \frac12$ (cf. \eqref{aal}). The associated
value of ${\tilde g}  = 2 m g $ is $\frac 34 \hbar^2$.
The heat kernel $K_E$ is determined in terms of the modified Bessel function $I_1$. Setting $u = 2 \omega T$, the partition function
\eqref{trR} reads $Z(u) = 1/(e^u-1)$.
As shown in \cite{LopesCardoso:2025kry}, a regularisation of the integral $ \int_{\mathbb{R}_0^+} (Z(u)/u) \, du $ yields the one-loop effective action for a free CFT of central charge $c=\frac32$ in global Euclidean $AdS_2$.

\subsection{Contribution from the constant path}

Let us return to the heat kernel expression \eqref{heatomx0} for
periodic paths. 
In the limit $\omega T \rightarrow 0$, 
the classical periodic path contributing to this 
expression is a constant path, as follows.

The equation of motion derived from the Euclidean Lagrangian \eqref{euclLag},
\bea
m {\ddot{x}} = - \frac{2g}{x^3} + m \omega^2 x \;,
\eea
admits the constant solution (we take $x_0 >0$)
\bea
x_0^4 = \frac{2g}{m \omega^2} \;.
\label{x04}
\eea
The classical action, when evaluated on this solution, gives
\bea
S_E (x_0) = \int_{t_i}^{t_f} L_E (x, \dot{x}) \, dt =  \sqrt{2 m g}  \; \omega T  \;\;\;,\;\;\; T = t_f - t_i \;.
\eea
Let us then consider the 
heat kernel $K_E$ expression \eqref{heatomx0} for periodic paths that start and end at $x_0$. The constant path contribution ${\rm exp} [- S_E (x_0) / \hbar]$ arising in the limit\footnote{Any other non-static configuration that extremises the Euclidean Lagrangian \eqref{euclLag} in the vanishing $\omega T$ limit will have a higher  on-shell value of the action than the vanishing  static configuration value, and hence will be sub-dominant in the path integral.} 
$\omega T \rightarrow 0$
can be extracted as follows. Namely, for large $z$, the modified Bessel function $I_a (z) $ behaves as (cf. eq. (2.13) in \cite{Dabholkar:2005dt})
\bea
I_{a}(z) = \left( \frac{1}{2 \pi z} \right)^{1/2} \, e^{ \left[ z - \frac{(4 a^2 -1)}{8 z} + {\cal O} \left( \frac{1}{z^2} \right) \right] } 
\;\;\;,\;\;\; a =  \frac12 \left( 1 + \frac{8 m g}{\hbar^2} \right)^{1/2} > 0 \;.
\label{asymhI3}
\eea
Thus, its exponent behaves as
 \bea
 z- \frac{(4 a^2 -1)}{8 z} = z - \frac{m g}{\hbar^2 z}  \;.
 \label{expzI}
 \eea
In \eqref{heatomx0}, the argument of the modified Bessel function $I_a$ is 
\bea
z = \frac{m \omega }{\hbar} 
\frac{ x^2 }{\sinh  (\omega T )} \;.
\label{valz}
\eea
 Using \eqref{expzI} and adding up 
  the exponential terms 
  in \eqref{heatomx0}, we obtain
\bea
- \frac{m \omega}{ \hbar} x^2  \coth  (\omega T)  + 
\frac{m \omega }{\hbar} 
\frac{ x^2 }{\sinh  (\omega T )} 
- \frac{ g \sinh  (\omega T )}{\hbar \omega x^2} \;.
\label{comb3}
\eea
On the constant solution \eqref{x04}, $z$ given in \eqref{valz} evaluates to
\bea
z = \frac{\sqrt{2 m g}}{\hbar \, \sinh(\omega T)} \;,
\label{zx0}
\eea
while the combination \eqref{comb3} takes the value
\bea
\frac{\sqrt{2 m g}}{\hbar \, \sinh(\omega T)} \left( 1 - \cosh (\omega T)  - \frac12  \sinh^2 (\omega T) \right) \;.
\eea
Expanding the latter around $\omega T \rightarrow 0$ gives
\bea
- \frac{\sqrt{2 m g}}{\hbar } \omega T + {\cal O} ((\omega T)^2) = - \frac{S_E (x_0)}{\hbar} + {\cal O} ((\omega T)^2 ) \;.
\eea
Thus, the classical action $S_E (x_0)$ emerges when taking the limit $\omega T \rightarrow 0$. In this limit, the expression for $z$ given in \eqref{zx0} becomes large and reduces to
\bea
z = \frac{\sqrt{2 m g}}{\hbar \, \omega T}  = \frac{m x_0^2}{\hbar T}\;,
\label{zcwt}
\eea
where we used \eqref{x04} to express $z$ in terms of $x_0$.

The 
heat kernel $K_E$ expression \eqref{heatomx0} for periodic paths that start and end at the point $x_0$ given by \eqref{x04} 
can be approximated by 
\bea
{\tilde K}_E (z,a) = 
\sqrt{z}
  \, e^{- z } \, I_a \left( z
 \right) \;,
\label{heatom3}
\eea
with $z$ given in \eqref{zcwt}. Namely, 
the expression for the heat kernel $K_E (x_0, x_0, T)$ 
differs from
${\tilde K}_E (z,a)$ by subleading corrections in $\omega T$. To lowest order in $\omega T$ we obtain 
\bea
\sqrt{\frac{\hbar T}{m}} \, K_E (x_0, x_0, T) &=& {\tilde K}_E (z,a) + c(z)\,
(\omega T)^2  + g(z) \, {\cal O} ( (\omega T)^{4}) \;, \nonumber\\
c(z) &=& - \frac16 \sqrt{z} \, e^{-z} \left( (1 + a + 2 z ) I_a (z) + z I_{a+1} (z) \right) \;.
\label{heatKexp}
\eea

\section{BPS black holes and the holographic DFF$_\omega$ model}\label{Embedding}

In this section we show how BPS black hole degeneracies can be extracted from a holographic CQM given by \eqref{euclLag} via the constant solution \eqref{x04} in the 
small $\omega T$ limit.

\subsection{Holographic DFF partition functions}

We focus on two classes of BPS black holes, namely $\tfrac12$ BPS small black holes in toroidally compactified
heterotic string theory \cite{Dabholkar:2005dt} and $\tfrac18$ BPS large black holes in toroidally compactified type II superstring theory \cite{Maldacena:1999bp}.
In both cases, an exact expression for the black hole microstate degeneracy $d(n)$ can be obtained in terms of a Rademacher series expansion. We wish to interpret the exact 
expression for $d(n)$ in terms of an appropriate holographic CQM.

We postulate the holographic CQM to be 
a model of DFF particles belonging to $s$ different species, each consisting of a very large number $N$ of DFF particles.
This CQM has an $SL(2, \mathbb{R})$ symmetry, a necessary requirement for being holographically dual to $AdS_2$. 
In order to quantise the system and obtain a discrete spectrum, in each of the $s$ sectors we couple  
the  $N$-particle DFF Hamiltonian to a simple harmonic oscillator \footnote{This is equivalent to choosing to work in the compact black hole/Rindler time from the bulk perspective.}, 
\bea {\mathcal H} = \sum_{i=1}^N \left(\frac{p_i^2}{2 m} + \frac{g}{x_i^2} + \frac{1}{2} m \omega^2 x_i^2\right) \;\;\;,\;\;\; g> 0 \;.
\label{caldff}
\eea 
Any such proposal for a holographic CQM must satisfy
two necessary conditions: 
\begin{enumerate}
    \item Derive the microscopic state degeneracy $d(n)$ as arising from a spectral degeneracy count \eqref{DFFcount} of the holographic DFF$_\omega$ model based on the generator $R$ (cf. \eqref{genR}).
    
    \item Explicitly demonstrate how the DFF$_\omega$ model couplings, $g$ and $\omega$, appear as parameters with fixed values in the microstate degeneracy formula for $d(n)$. 
    
\end{enumerate}
We will show that the above postulate indeed satisfies these conditions in the case of certain classes of  BPS black holes, and specify the coupling constants of the theory based on these criteria. 
We will take $s \geq 1$ to be the number of species of these particles\footnote{We are motivated to define a species number, with the foreknowledge that string models of microscopic black hole degeneracy such as that for $\frac{1}{2}$ BPS black holes in $\mathcal{N}=4$ heterotic toroidal compactifications involve multiple non-interacting scalar fields (24 in this case). We will see that the species number does get fixed in degeneracy matching between the microscopic and holographic formulas.}, and assume that the species are mutually non-interacting.   Following \cite{Lechtenfeld:2015wka}, the counting formula for the state degeneracy $p_M$ at energy level $M$, for each species, in the ``thermodynamic" limit of large number of  DFF$_\omega$ particles, is simply the number of partitions 
of $M$ into sums of positive integers, without regard to the order of the summands\footnote{\label{1} If $N$ is the number of particles, then the counting formula is $p_N(M)$, the number of partitions of $M$ into $N$ parts. For $N \rightarrow +\infty$ \cite{hardy1918asymptotic}, this simply becomes $p_M$, the unconstrained number of partitions of $M$, with $\displaystyle{\lim_{N\rightarrow +\infty}}\frac{p_N(M)}{p_M} \approx 1 - e^{-{\cal O}(\frac{N}{\sqrt{M}})}$.} 
and hence, the corresponding generating function for the degeneracy  is the Euler partition function expressed as a function of the Euclidean time periodicity $\beta$, complexified to $\tau = \tau_1 + i \beta/2 \pi$, 
\begin{equation}
    \sum_{M \in \mathbb{N}_0} p_M \, q^M = \prod_{k=1}^{\infty} \frac{1}{(1-q^k)}= \frac{q^{\frac{1}{24}}}{\eta(q)}  \;\;\;,\;\;\; q= e^{2\pi i \tau} \;,
\end{equation}
where $\eta$ denotes the Dedekind $\eta$ function. 
Thus the total degeneracy, taking into account all the species, is extracted by an inverse Laplace transform, 
\begin{equation}\label{DFFcount}
    p_M = 
       \int_{\mathcal{C}}
       d\tau \frac{1}{\eta^s (e^{2 \pi i \tau})}e^{-2 \pi i \tau (M-\frac{s}{24})} \;,
\end{equation}
where the integration contour $\mathcal{C}$ is chosen to be the one appropriate for the circle method calculation of its Rademacher series expansion, as shown below in \eqref{dnsmall}.

The large $M$ asymptotic degeneracy is encoded in the modified Bessel function of the first kind $I_a (z)$, whose index $a$ is related to $s$ by $a = 1 + s/2$, and reads
\begin{equation}
    p_{M}  \approx e^{z - (\frac{s + 3}{4}) \log\, ( M s)},
    \label{plogs}
\end{equation}
where $z$ is
\bea
z = 2 \pi \sqrt{\frac{M s}{6}} \;.
\label{zms}
\eea

We will now show how the DFF proposal does satisfy the two necessary conditions for a holographic CQM in the ${\cal N}=4$ and ${\cal N}=8$ cases, by firstly identifying the value of $s$ that enables us to reproduce microscopic degeneracies from the holographic CQM picture, and secondly explicitly demonstrate how the DFF$_\omega$ couplings ($g, \omega$) appear as parameters in the microstate degeneracy formula for $d(n)$.

\subsubsection{$\frac12$ BPS small black holes}

We consider small $\frac12$ BPS black holes in heterotic string theory compactified on a six-torus \cite{Dabholkar:2005dt}. The microstate degeneracies $d(n)$ depend on an integer $n$, which is a bilinear in the charges carried by the black hole. The generating function of the degeneracies $d(n)$ is $1/\eta^{24} (q)$, where $\eta$ denotes the Dedekind eta function, and where $\eta^{24} (q) = q \prod_{n=1}^{\infty} (1 - q^n)^{24}$. 
An exact expression for $d(n)$ can be given in terms of a Rademacher series expansion. It reads (see, for instance, eq. (C.4) in \cite{LopesCardoso:2021aem}), 
\bea
d(n) &=& \int_{\mathcal C} d\tau \frac{1}{\eta^{24}(\tau)} e^{-2\pi i \tau n} \nonumber\\
&=& \sum_{\gamma =1}^{\infty} \, 
\sum_{\substack{0\leq -\delta<\gamma\\ (-\delta,\gamma)=1}}
\int_{\bigcirc_{\delta,\gamma}} d \tau \frac{1}{\eta^{24}(\tau)} e^{-2\pi i \tau n}\nonumber\\
&=&2 \pi \displaystyle \sum_{\gamma =1}^{\infty} \frac{1}{\gamma \, n^{13/2} } \,
{\rm KL}(n, -1, \gamma)  \, {I}_{13} (z) \;\;,\;\; z= \frac{4 \pi \sqrt{n}}{\gamma} \;, 
\label{dnsmall}
\eea
where $n \gg 1$ denotes a bilinear combination of charges carried by the BPS black hole. Here, $I_{13}$ denotes the modified Bessel function of the first kind with index $a = 13$, and
${\rm KL}(n, -1, \gamma)$ denotes the classical Kloosterman sum given by
(see, for instance, eq. (C.29) in \cite{LopesCardoso:2021aem}), 
\bea
{\rm KL} (n, -1, \gamma) =  \sum_{\substack{0\leq-\delta<\gamma\\ \alpha\delta = 1 \text{ mod } \gamma}}e^{2\pi i\left(- \frac{\alpha}{\gamma}+ n\frac{\delta}{\gamma} \right)}.
\label{classKloo}
\eea
In the above computation of the Rademacher series expansion  for $d(n)$, the integration contour $\mathcal{C}$, which
is any contour in the
complex upper half plane that starts at some point $\tau_0$ and ends at $\tau_0  + 1$, is deformed into a new integration contour that is the union \cite{rademacher1937convergent} of all the Ford circles $\bigcirc_{\delta,\gamma}$, where $-\delta/\gamma$ define all rationals on the unit interval $[0,1)$. 

The expression for the microstate degeneracies $d(n)$ is equal to $p_{n-1}$ in the DFF$_\omega$ degeneracy formula \eqref{DFFcount} for $s=24$. This satisfies the first of the two necessary conditions mentioned above. To demonstrate the relation between the DFF$_\omega$ couplings and the microstate degeneracy formula parameters, we return to the expression for $d(n)$ as a Rademacher series \eqref{dnsmall}.  

The Rademacher series expansion expresses the microscopic degeneracy $d(n)$ in terms of a sum over the rationals in $[0,1)$, with each summand consisting of three types of contributions -- the polar terms which are microscopic degeneracies $d(n)$ for all $n<0$,\footnote{In this case, this is simply a single number $d(-1)$.} a modified Bessel function of the first kind with index fixed uniquely in terms of the modular weight of the generating function, and phases (corresponding to the rational number associated with each summand) encoded in the Kloosterman sum \eqref{classKloo}. Hence, apart from the polar term, the only parameters governing the expansion are the index $a$ and the argument $z$ of the modified Bessel function. We need to show how they arise from the DFF$_\omega$ parameters. Note that the DFF model potential is an inverse square potential, with the simple harmonic oscillator tacked on for the purpose of obtaining a discrete spectrum upon quantisation, cf. \eqref{euclLag}. Hence, the two couplings of relevance are $g$ and $\omega$. We have already shown in \eqref{indamg} the relation between $a$ and $g$. We will relate the frequency $\omega$ to the argument $z$ in \eqref{zcom}  below.

We use the relation \eqref{heatom3} to express $I _{13}$ in \eqref{dnsmall} in terms of the heat kernel
${\tilde K}_E$, 
\bea
I_{13}(z) = z^{-1/2} \, e^z \, {\tilde K}_E (z, a=13) \;.
\label{besskap}
\eea
Substituting this into \eqref{dnsmall} gives
\bea
d(n) = \frac{1}{ \sqrt{\pi}  } \displaystyle \sum_{\gamma =1}^{\infty} \frac{1}{\sqrt{\gamma }}  \,
{\rm KL}(n, -1, \gamma)  \, e^{z- \frac{27}{4} \log n}  \,  {\tilde K}_E(z, a=13) \;\;,\;\; z= \frac{4 \pi \sqrt{n}}{\gamma} \;.
\label{dzn}
\eea
Using the expression \eqref{indamg} for the index $a$,
 \bea
 \frac{2 m g}{\hbar^2}  = 
a^2 - \frac14   \;,
\label{mga}
\eea
 we infer that the combination $m g/\hbar^2$ cannot depend on black hole data such as charges, since the index $a$ is a real number that takes a fixed value in a given theory. Using \eqref{mga}, the value of $z$ given in \eqref{zcwt} can be expressed as
\bea
z = \frac{\sqrt{a^2 - \frac14}}{\omega T} \;.
\label{zcom}
\eea
Since, in the limit $\omega T \rightarrow 0$, $z$ takes a large value,  it is consistent to identify $z$ 
with the large charge leading order contribution to the BPS black hole entropy, 
\bea
z =  \frac{4 \pi \sqrt{n}}{\gamma} \;,
\label{zat}
\eea
where we recall that $n \gg 1$ is the bilinear function of black hole charges that appears in the BPS black hole, with $\gamma \in \mathbb{N}$. By equating  \eqref{zcom} and \eqref{zat}, we infer 
\bea
\omega T  =  \frac{\gamma \sqrt{a^2 - \frac14}}{ 4 \pi \sqrt{n} } \;,
\label{wtzrel}
\eea
where we recall that $a$ is a fixed number that is independent of $\gamma$ and of the charge bilinear $n$.

{From} the supersymmetric gravitational path integral analysis \cite{Murthy:2009dq,Sen:2009gy}, we see that it is consistent to identify $\gamma$ with the periodicity $T$ of Euclidean time.  We may therefore set $T$ to $T=\gamma$, in which case \eqref{wtzrel}
becomes an equation expressing $\omega$ in terms of  $a, n$.
Consequently, the equation \eqref{wtzrel} states that the holographic CQM that corresponds to the near-horizon $AdS_2$ of a $\frac{1}{2}$ BPS black hole with a charge invariant $n$  must have a specific coupling $\omega$ defined as above, with the index $\gamma$ characterising each sector of the Rademacher expansion in the degeneracy formula encoded in $T$. 

Thus, we have related the combinations $m g/\hbar^2$ and $\omega $ to the data $a, n$ 
that enters in the microscopic state counting, where the limit $n\rightarrow + \infty$ is aligned with $\omega \rightarrow 0$.

Recall that the heat kernel $K_E$ for periodic paths that start and end at the point $x_0$ given by \eqref{x04}, reduces to $ {\tilde K}_E$ in the limit of small $\omega $,  wherein $x_0  = \sqrt{\hbar  z /m}$. Then using the asymptotic expansion of $I_{13}$ given in eq. (2.13) of \cite{Dabholkar:2005dt}, we infer from \eqref{besskap} above that 
\bea
{\tilde K_E}(z, a) 
= \frac{1}{\sqrt{2 \pi}} \left( 
 1 - \frac{(4 a^2 -1)}{8 z} + 
\frac{(4 a^2 -1)(4 a^2 -3^2)}{2! (8z)^2} - \dots \right)  \;,
\label{tKE}
 \eea
 where here $a=13$. Recalling that the value of $ \gamma \, z = 4 \pi \sqrt{n}  $ equals the leading order BPS entropy of the small black hole, we see that the heat kernel ${\tilde K_E}(z, a=13)$ 
computes all the power law suppressed corrections to BPS black hole entropy. 
The limit $\omega \rightarrow 0$  is consistent with the holographic DFF$_\omega$ model being postulated as an effective description of the `true' CQM (cf. \eqref{caldff}), with the requirement that it yields the exact degeneracy of ground states in the global $AdS_2$ frame. We show that this requirement is indeed met, as follows.

A state at energy level $M$ in the holographic model gets mapped to the ground state in global $AdS_2$. This can be seen as follows. The generator $R$ in \eqref{genR} is the generator of translations in compact time, $R = - i d/d \varphi$, and hence, the associated Euclidean $AdS_2$ metric is the one written in terms of compact black hole/Rindler time, 
\bea
ds_2^2 = a^2 \left( \sinh^2 \rho \, d \varphi^2 + d \rho^2 \right) \;\;\;,\;\;\; \varphi \equiv \varphi + 2 \pi \;.
\label{rindads}
\eea
The metric of Euclidean $AdS_2$ metric in global time, on the other hand, is given by
\bea
ds_2^2 = \frac{a^2}{\sin^2 \sigma} \left( d \tau^2 + d \sigma^2 \right) \;\;\;,\;\;\; - \pi < \sigma < 0  \;.
\label{gloads}
\eea
The segment  $\rho = \infty, - \pi/2 < \varphi < \pi/2$ of the $AdS_2$ boundary  gets mapped to $\sigma = 0$  \cite{Sen:2011cn}, 
with $\varphi$ and $\tau$ being related by
\bea
\tan \frac{i \tau}{2} = \tanh \frac{i \varphi}{2} 
\;,
\eea
which results in 
\bea
\frac{\partial}{\partial \tau }= \frac{\cos^2 \frac{\varphi}{2}}{\cosh^2 \frac{\tau}{2}} \, \frac{\partial}{\partial \varphi } \:.
\eea
Note that the conversion factor is non-singular. 
Taking into account the warp factor of the metric \eqref{gloads}, it follows
that close to the boundary $\sigma = \epsilon \rightarrow 0$, a state at energy level $M$ in thermal time $\varphi$ gets mapped to a state
with energy proportional to $\sin \epsilon \,  M \rightarrow 0$ in global time $\tau$, and hence to the ground state of $AdS_2$.

Thus, we have accurately reproduced the microstate degeneracy formula for $\frac12$ BPS black holes \eqref{dnsmall} 
from a holographic DFF$_\omega$ model picture through the parameter map \eqref{mga}, \eqref{wtzrel}, with $T = \gamma$.

\subsubsection{$\frac18$ BPS large black holes}

Type II string theory compactified on a six-torus $T^6$ admits large $\frac18$ BPS black holes, whose microstate degeneracies $d(\Delta)$ can be conveniently computed in particular duality frames \cite{Maldacena:1999bp,Shih:2005qf}, in which the charges carried by the black hole have an explicit description in terms of 
charges in the microscopic string theory. The simplest realization of such BPS black holes are black holes that carry 5 charges.
In the type IIB description on $T^6 = T^4 \times S^1 \times {\tilde S}^1$, 
consists 
in choosing charges that correspond to one D1 brane wrapped on $S^1$, one D5 brane wrapped on $T^4 \times S^1$, one unit of Kaluza-Klein monopole charge associated with ${\tilde S}^1$, $n$ units of Kaluza-Klein momentum on $S^1$ and $\ell$ units of Kaluza-Klein momentum on ${\tilde S}^1$. For this black hole realization, the
quartic U-duality invariant charge combination $\Delta$ reads
$\Delta = 4 n - \ell^2$.  

The generating function for the microstate degeneracies $d(\Delta)$ of the above BPS black hole is given by the weak Jacobi form of weight $-2$ and index $1$,
\bea
\varphi_{-2,1} (\tau,z) = \frac{ \vartheta_1^2 (\tau,z)}{\eta^{6}(\tau)} = \sum_{n, \ell \in \mathbb{Z}} c(n, \ell) \, q^n \, y^{\ell} \;\;\;,\;\;\; q= e^{2 \pi i \tau} \;\;,\;\; y = e^{2 \pi i z} \;,
\eea
with $c(n, \ell) = C_{\ell} (\Delta)$, where $C_{\ell} (\Delta)$ depends only on $\ell = \Delta \mod 2$.
The microstates degeneracy $d(\Delta)$ is given in terms of $C_{\ell} (\Delta) $ as
$d(\Delta) = (-1)^{\Delta +1} C_{\ell}(\Delta) $. The exact expression for $C_{\ell}(\Delta)$ with $\Delta \geq 1$ is 
given in terms of a Rademacher series expansion and reads \cite{Dijkgraaf:2000fq,Manschot:2007ha,Dabholkar:2014ema,Murthy:2015zzy} 
\bea
C_{\ell} (\Delta) = 2 \pi \left(\frac{\pi}{2} \right)^{7/2}  \displaystyle \sum_{\gamma =1}^{\infty} \frac{1}{\gamma^{9/2}  } \,
{\rm KL}(\Delta, {\ell})  \, {\hat I}_{\frac72} (z) \;\;,\;\; z= \frac{\pi \sqrt{\Delta}}{\gamma} \;,
\label{Cdel}
\eea
where $\hat{I}_{\frac72}$ denotes the hatted modified Bessel function of first kind with index $\frac72$,
$I_{7/2} (z) = \left( \frac{z}{2}\right)^{7/2} \, {\hat I}_{7/2} (z) $.
The Kloosterman sum $\rm KL$ reads (cf. eq. (2.19) in \cite{Dabholkar:2014ema})
\bea
{\rm KL} (\Delta, \nu) =  \sum_{\substack{0\leq-\delta<\gamma\\ \alpha\delta = 1 \text{ mod } \gamma}} e^{2\pi i\left(\frac{\Delta}{4} \,\frac{\delta}{\gamma} \right)} \, M^{-1}_{\nu, 1} \, e^{2\pi i\left(- \frac14 \, \frac{\alpha}{\gamma} \right)} \;\;\;,\;\;\; \text{with} \,\, \nu = \Delta \mod 2 \;,
\label{Kloon8}
\eea
where $M_{\mu, \nu}$ is called the multiplier system. We refer to eq. (2.15) in \cite{Gomes:2017bpi} for an explicit expression for this multiplyer system.

Now we proceed to analyse the various contributions in \eqref{Cdel} from the point of view of the boundary DFF$_\omega$ model, as in the case of small BPS black holes. The hatted Bessel function ${\hat I}_a$ is related
to the Bessel function $I_a$ by $I_a (z) = \left( \frac{z}{2}\right)^a \, {\hat I}_a (z) $. Hence, using
\eqref{heatom3} to express ${\hat I}_{\frac72}$ in terms of ${\tilde K}_E$, 
\bea
{\hat I}_{\frac72} (z) = 2^{7/2} \, \frac{e^z}{z^4} \, {\tilde K}_E (z, a = \tfrac72 ) \;,
\eea
we obtain
\bea
C_{\ell} (\Delta) =  2 \sqrt{\pi} \,  \displaystyle \sum_{\gamma =1}^{\infty} \,
\frac{1}{\sqrt{\gamma} } \,
{\rm KL}(\Delta, \ell)  \, e^{z- 2 \log \Delta}  \, {\tilde K_E} (z, a=\tfrac72)\;\;,\;\; z= \frac{ \pi \sqrt{\Delta}}{\gamma} \;,
\label{Cdel2}
\eea
where we recall that ${\tilde K_E} (z, a)$ has the asymptotic expansion \eqref{tKE} which in the present case terminates, since $a = 7/2$ is half integer valued. Thus, as in the ${\cal N}=4$ case, the heat kernel ${\tilde K_E}$ associated with a DFF$_\omega$ model on the boundary of $AdS_2$ accounts for all the power law
suppressed corrections to BPS black hole entropy. The value of $a$ defines the weight of the corresponding partition function $1/\eta^s (\tau)$ (cf. \eqref{as}), setting the number of species of the DFF$_\omega$ particles to be $s=5$, thereby defining the corresponding holographic generating function of microscopic degeneracies to be $1/\eta^5 (\tau)$. From \eqref{plogs} and \eqref{zms}, the degeneracies generated by this partition function are easily seen to asymptotically yield the expected
leading term in the BPS entropy and the logarithmic correction term in \eqref{Cdel2},
$e^{z - 2  \log \Delta}$, where $z$ is the argument of the modified Bessel function of index $a = 1 + \frac{s}{2}=\frac{7}{2}$, appearing in the Rademacher series expansion \eqref{Cdel}. 

Next, let us turn to the Kloosterman sum \eqref{Kloon8} in \eqref{Cdel2}. Unlike in the ${\cal N}=4$ case, where the exact microstate counting formula itself was a weakly holomorphic modular form, in this case, we can obtain neither the phases above nor the multiplier system $M^{-1}_{\nu, 1}$ or the factor of $1/\sqrt{\gamma}$ from the DFF$_\omega$ model point of view, reflecting the fact that the exact microstate counting function is a Jacobi form, which carries additional structure compared to a modular form, and is subsequently captured with less fidelity by a holographic DFF$_\omega$ model and the modular form associated with it.

\subsubsection{Assessment}

In each of the above cases, the exact microscopic counting function is encoded in terms of a modular or weak Jacobi form of negative weight, whose Fourier coefficients are expressed in terms of a modified Bessel integral of the first kind, which allows the holographic DFF$_\omega$  picture to accurately model the asymptotic leading, logarithmic as well as sub-leading power law contributions to the microstate degeneracy. In contrast,  $\frac14$ BPS large black holes in ${\cal N}=4$ toroidal heterotic compactifications or $\frac12$ BPS large black holes in the ${\cal N}=2$ STU or FHSV models have microscopic counting functions expressed in terms of Siegel modular forms \cite{Dijkgraaf:1996it,Cardoso:2019avb}, and the corresponding Fourier coefficients are no longer expressed solely in terms of a modified Bessel function of the first kind.  This is correlated to the numerical prefactors of the log terms in both these cases \cite{Sen:2012kpz} (namely zero for ${\cal N}=4$, and positive for the STU model) not being amenable to being encoded in a DFF$_\omega$ description.

\subsection{Values of ${\tilde \alpha}, a, {\tilde g}, s$}

We have shown above that a judicious choice of the number $s$ of species of DFF$_\omega$ particles correctly reproduces the leading large charge BPS entropy as well as the logarithmic correction term  thereof. We have also shown
that in the limit $\omega  \rightarrow 0$, the heat kernel ${\tilde K_E}$ of the holographic DFF$_\omega$ model living at the boundary of the BPS black hole near-horizon Euclidean $AdS_2$ space-time reproduces the power law suppressed corrections to its entropy, for a fixed number of species $s$ of DFF$_\omega$ particles. This determines the corresponding holographic generating function to be $1/\eta^s (\tau)$. This function equals
the microscopic generating function for $\tfrac12$ BPS small black holes, and generates all contributions to the 
microstate degeneracy of $\tfrac18$ BPS large black holes except for the Kloosterman phases and the factor of $1/\sqrt{\gamma}$.
 Different types of BPS black holes require different values of the parameters in the holographic DFF$_\omega$ model on the boundary and of the number of species $s$ of the DFF$_\omega$ particles. This is summarized in Table \ref{tab1}, where we have also included the case of global $AdS_2$ with a free CFT of central charge $c = \frac32$ and $s=0$, as
discussed at the end of Subsection \ref{sec:pf}. In Table \ref{tab1} we make use of the relations \eqref{tga} and \eqref{aal}.

\vskip 3mm

\begin{table}[H]
\begin{center}
\begin{tabular}{|c | c | c | c | c|} 
\hline
Type & ${\tilde \alpha}$ & Bessel index $a$  & coupling ${\tilde g}/\hbar^2$ & species $s$ \\
\hline
Global $AdS_2$ ($c=\frac32$) & 1/2 & 1 & 3/4 & --\\ 
\hline 
$\tfrac12$ BPS small black hole & 25/2   & 13 & 675/4 & 24 \\
\hline
$\tfrac18$ BPS large black hole & 3 & 7/2 & 12 & 5 \\
\hline
\end{tabular}
\caption{Bessel index $a = {\tilde a} = {\tilde \alpha} + \frac12$. Coupling
${\tilde g} = {\tilde \alpha} ({\tilde \alpha} + 1) \hbar^2 $. \label{tab1}}
\end{center}
\end{table}

Note that there is a clear separation between the ${\cal N}=8$ case and the other two cases. Namely, in the ${\cal N}=8$ case, the Bessel index $a$ is half-integer valued, whereas it is integer valued in the $AdS_2$ case and in the small black hole case. This implies that the expansion
\eqref{tKE} terminates in the ${\cal N}=8$ case.

 \section{Conclusions: Microscopic BPS counting from holographic boundary DFF$_\omega$ models}
 
 In this note, we have taken steps towards providing a physical interpretation of the exact microstate degeneracies of 4D $\tfrac12$ BPS small black holes in a ${\cal N}=4$ heterotic and $\tfrac18$ BPS large black holes in the toroidal ${\cal N}=8$ type II superstring compactifications, in terms of the holographically dual boundary conformal quantum mechanics (CQM) associated with the black hole near-horizon $AdS_2$ space-times. The length scale ($AdS$ radius) of the near-horizon geometry is fixed by the charge bilinear $n$ of the black hole. Hence, for a fixed $n$, we have a fixed near-horizon geometry and its corresponding holographic setup.   The black hole microstates degeneracies in the above  models are known to be generated by a weakly holomorphic weight $-12$ modular form, $1/{\eta^{24}(\tau)}$, and a weak weight $-2$ Jacobi form, $\vartheta_1^2(\tau,z)/\eta^6(\tau)$, respectively, with the corresponding Fourier and Fourier-Jacobi coefficients being computed as a convergent Rademacher expansion series, via the Hardy-Littlewood circle method.  This series is organised in terms of fractions $0 \leq - \delta/\gamma< 1$ spanning $\mathbb{Q}\bigcap [0,1)$, with each $\gamma$ sector encoding the degeneracy contribution from a $\gamma$ orbifold of  $AdS_2 \times S^2$ \cite{Banerjee:2008ky,Murthy:2009dq,Dabholkar:2014ema}. Each summand in the Rademacher series expansion in a given $\gamma$ sector contains a modified Bessel function of the first kind, whose index is determined by the weight of the generating function, as well as a Kloosterman sum constituted of phase factors. This series expansion carrying distinct types of terms in each summand turns out to be optimal for establishing the relation between the DFF$_\omega$ model couplings and the various terms in the microstate degeneracy formula. 
 \par 
 Motivated by a modeling of the BPS black holes under consideration as a collection of Calogero particles moving in the near-horizon $AdS_2$ space-times, we postulate that the dual boundary CQM has a universal de Alfaro-Fubini-Furlan sector constituted of $s$ species, each possessing $N \gg 1$ particles of 
  mass $m$ with $m/N\rightarrow 0$, independent of the specifics of the theory under consideration. The corresponding generating function of spectral degeneracies is $1/\eta^{s} (\tau)$. In the ${\cal N}=4$ case, $s=24$ yields the generating function $1/\eta^{24}(\tau)$ to be precisely the exact microscopic generating function, and this choice also therefore yields the precise Kloosterman sum \eqref{classKloo} along with the factor $1/\sqrt{\gamma}$, with the identification of $\omega$
 as inversely proportional to $\sqrt{n}$, cf. \eqref{wtzrel} with $T=\gamma$.
 We also showed that the corresponding DFF$_\omega$ heat kernel trace  ${\tilde K_E}$, corresponding to $ \omega \rightarrow 0$, encodes the sub-leading power law suppressed corrections to the microstate degeneracy in each Rademacher summand, with the index of the modified Bessel function in the Rademacher series expansion fixed in terms of the DFF coupling $g$. In the ${\cal N}=8$ case, the microscopic counting function is a Jacobi form with a more complicated Kloosterman structure. Here, the value of $s=5$ captures all terms in each Rademacher summand except the Kloosterman phases and the factor of $1/\sqrt{\gamma}$.  We note that the limit $\omega  \rightarrow 0$   is consistent with the holographic DFF$_\omega$ model being postulated as an effective description of the `true' CQM (cf. \eqref{caldff}), with the requirement that it yields the exact degeneracy of ground states in the global $AdS_2$ frame.

Our results here are only the first steps in integrating two distinct microscopic descriptions of BPS black holes -- the holographic CFT construction of gravity and a microscopic string compactification picture leading to exact state counting formulae. Several open problems offer themselves immediately -- rigorous holographic interpretations of the Kloosterman sums in ${\cal N}=8$, as well as interpreting the Rademacher series expansion of the coefficients of the Siegel modular form that generates $\tfrac14$ BPS large black holes in ${\cal N}=4$ \cite{Ferrari:2017msn,LopesCardoso:2021aem}, which is far more intricate than that of a modular or Jacobi form. Answers to these questions can provide a blueprint to integrate two of the most successful approaches to modeling quantum gravity -- an algebraic geometric picture of string compactifications with associated enumerative invariants underlying black hole degeneracies, and  a conformal field theoretic picture of quantum gravity.   We hope to pursue and report on these questions in future publications.


\section*{Acknowledgements}
We would like to thank Jos\'e Barb\'on, Roberto Emparan, Robert de Mello Koch and Thomas Mertens for helpful discussions.
Research partially funded by Funda\c{c}\~ao para a Ci\^encia e Tecnologia (FCT), Portugal, through grant No. UID/4459/2025
and through CEMS.UL, project UID/04561/2025.

\appendix

\section{Path integral representation of the heat kernel in quantum mechanics  \label{sec:heat} }

We review the path integral representation of the heat kernel $K$ in quantum mechanics following \cite{takh}.

Consider the Schr\"odinger equation for a quantum particle in one dimension, 
\bea
i \hbar \partial_t \psi =  H \psi \;,
\eea
where
\bea
H = 
- \frac{\hbar^2}{2m } \partial_x^2 + V(x) \;.
\eea
The Cauchy problem 
\bea
i \hbar \partial_t \psi (t,x) = 
\left(  - \frac{\hbar^2}{2m } \partial_x^2 + V(x)  \right) \psi (t,x) \;\;\;,\;\;\; \psi (t,x) \vert_{t =0} = \psi_0 (x) \;,
\eea
has a fundamental solution: a function $K(x, x', t )$, which satisfies
\bea
i \hbar \partial_t  K(x, x', t)  &=& 
\left( - \frac{\hbar^2}{2m } \partial_x^2 + V(x)   \right) K(x, x', t) \;, \nonumber\\
K(x, x', t=0) &=& \delta (x -x') \;,
\eea
in a distributional sense.
The solution to the Cauchy problem can then be formally written as
\bea
\psi (t,x) = \int_{\mathbb{R}}  K(x,x',t)  \, \psi_0 (x') d x' \;.
\eea

Let us consider the case when $H$ has a pure point spectrum, i.e. the spectrum is discrete and the Hilbert space has an 
orthonormal basis $\{\psi_n\}, n \in \mathbb{N}_0$, with eigenvalues $E_0 \leq E_1 \leq \dots \leq E_n \leq \dots$ of finite multiplicity,
\bea
H \psi_n (x) = E_n \psi_n (x) \;.
\eea
Then, $K$ has the eigenfunction representation
\bea
 K(x, x', t)  = \sum_{n\in \mathbb{N}_0} \, e^{- i \frac{E_n}{\hbar} t } \, \psi_n (x) \, {\bar \psi}_n (x') \;,
 \eea
 which converges in a distributional sense. Using $\psi_n(x) = \langle x | \psi_n  \rangle$, $ e^{- i \frac{ t H }{\hbar} } \, |\psi_n \rangle  = e^{- i \frac{E_n}{\hbar} t } \,| \psi_n \rangle$, 
 $ {\rm id} = \sum_n  |\psi_n \rangle \langle \psi_n |$, $K$ 
  becomes
 \bea
 K(x, x', t)  = \langle x | e^{- i \frac{ t H }{\hbar}  }  | x' \rangle \;\;\;,\;\;\; H  = - \frac{\hbar^2}{2m } \partial_x^2 + V(x) \;.
 \eea
 Denoting $K(x, x', t) = K(x, t; x', 0)$,  we have, 
 more generally,
  \bea
 K(x, t ; x', t')  = \sum_{n\in \mathbb{N}_0} \, e^{- i \frac{E_n}{\hbar} T } \, \psi_n (x) \, {\bar \psi}_n (x') \;\;\;,\;\;\; T = t  - t' \;.
 \eea

The heat kernel $K$ also has a representation in terms of a  Feynman path integral in configuration space, 
\bea
K(y, t_f; x, t_i) =  \langle y | e^{- i (t_f - t_i) \frac{  H }{\hbar}  }  | x \rangle = 
\int_{x(t_i) = x , \,  x(t_f) = y}  {\cal D} x(t) \, e^{\frac{i}{\hbar} S ( \gamma ) } \;,
\eea
where $S(\gamma)$ is the action functional evaluated on a path $\gamma$ between $(t_i, x)$ and $(t_f, y)$ (with $t_f > t_i$),
\bea
 S ( \gamma )  = \int_{t_i}^{t_f} L (x, \dot{x}) \, dt\;\;\;,\;\;\;
L(x, \dot{x}) = \frac12 m \dot{x}^2 - V(x) \;.
\eea

The Euclidean version of $K$, obtained by performing the analytic continuation $t \mapsto - i t $, reads
\bea
K_E(y, t_f ; x, t_i)  = \langle y | e^{-  \frac{ (t_f - t_i)  H }{\hbar}  }  | x \rangle = 
\int_{x(t_i) = x , \,  x(t_f) = y} {\cal D} x(t) \, e^{ - \frac{S_E ( \gamma )}{\hbar} } \;,
 \eea
where
\bea
S_E ( \gamma )  = \int_{t_i}^{t_f} L_E (x, \dot{x}) \, dt\;\;\;,\;\;\;
L_E(x, \dot{x}) = \frac12 m \dot{x}^2 + V(x) \; .
\label{heatpath}
\eea
The quantum mechanics partition function $Z(t) = {\rm Tr} e^{- \frac{t}{\hbar} H  }$ is obtained from $K_E(x,x,t) $,
\bea
K_E(x,x,t) =  \langle x | e^{-  \frac{ t H }{\hbar}  }  | x \rangle =  \sum_{n\in \mathbb{N}_0} \, e^{- \frac{E_n}{\hbar} t } \, \psi_n (x) \, {\bar \psi}_n (x) \;,
\eea
by
\bea
Z(t) = 
\int_{\mathbb{R}}  K_E(x,x,t) \,  dx = \sum_{n\in \mathbb{N}_0} \, e^{- \frac{E_n}{\hbar} t }  \;.
\label{heatZ}
\eea
In the Euclidean path integral formalism, the partition function $Z(\beta) $ is represented by a path integral over periodic orbits, 
i.e. paths $\gamma$  that start and end at the same point
$x_i (t_i=0) = x_f( t_f = \beta) = q_0$, 
\bea
\label{Zbe}
Z(\beta) = \int_{\mathbb{R}} dq_0 \left[  \int_{ x_i (t_i=0) = x_f( t_f = \beta) = q_0}   {\cal D} x(t) \, e^{- \frac{S_E ( \gamma ) }{ \hbar}}  \right]
\, ,\, S_E ( \gamma )  = \int_{0}^{\beta} L_E (x, \dot{x}) \, dt .
\nonumber\\
\eea


\section{Limiting cases of the heat kernel $K_E$ \label{sec:lim}}

The Euclidean heat kernel $K_E$ for the model described by the Euclidean Lagrangian \eqref{euclLag} is given by \cite{Khandekar}
\bea
 \label{heatom4}
K_E (y, t_f; x, t_i) = \frac{m \omega }{\hbar} \frac{\sqrt{ x y }}{ \sinh  \omega (t_f - t_i) }  \; e^{- \frac{m \omega}{2 \hbar} ( x^2 + y^2)  \coth  \omega (t_f - t_i)} \, I_a \left( \frac{m \omega }{\hbar} 
\frac{ x y }{\sinh  \omega (t_f - t_i)} \right) \;, \nonumber\\
\eea
where the expression for the index $a$ is given in \eqref{indamg}.
 In the following, we discuss various limiting cases of this expression, to recover well known formulae for propagators.

First, consider setting 
$\omega =0$, in which case we obtain the Euclidean heat kernel for the DFF model described by the Hamiltonian $H$ given in \eqref{HDFF},
\bea
K_E (y, t_f; x, t_i) = \frac{m }{\hbar} \frac{\sqrt{ x y }}{ (t_f - t_i) }  \; e^{- \frac{m}{2 \hbar (t_f - t_i) } ( x^2 + y^2)  } \, I_a \left( \frac{m  }{\hbar} 
\frac{ x y }{(t_f - t_i)} \right) \;.
\label{heatdff}
\eea
Defining
\bea
\beta = t_f - t_i \;,
\eea
setting $2 m = \hbar = 1$ and using \eqref{aal}, we obtain
\bea
K_E (y, t_f; x, t_i) =  \frac{\sqrt{ x y }}{ 2 \beta }  \; e^{- \frac{ ( x^2 + y^2) }{4 \beta} } \, I_{{\tilde \alpha} + \frac12} \left( 
\frac{ x y }{ 2 \beta } \right) \;,
\label{heatdff2}
\eea
in agreement with the result given in eq. (34) of \cite{Akhoury:1984pt}.

Next, consider setting $g=0$ in \eqref{heatom4}, which results in $a = 1/2$ (cf. \eqref{indamg}).
The resulting heat kernel is the one for the harmonic oscillator,
\bea
\label{heatharm}
K_E (y,t_f; x, t_i) = \frac{m \omega }{\hbar} \frac{\sqrt{ x y }}{ \sinh  \omega (t_f - t_i) }  \; e^{- \frac{m \omega}{2 \hbar} ( x^2 + y^2)  \coth  \omega (t_f - t_i)} \, I_{\frac12} \left( \frac{m \omega }{\hbar} 
\frac{ x y }{\sinh  \omega (t_f - t_i)} \right) \;,
\nonumber\\
\eea
where (cf. \eqref{asymhI3})
\bea
I_{\frac12} (z) = \left( \frac{1}{2 \pi z} \right)^{1/2} \, e^z \;.
\label{sol1}
\eea
Note that this modified Bessel function is singular at $z=0$. We get
\bea
K_E (y, t_f; x, t_i) = \sqrt{\frac{m \omega }{ 2 \pi \hbar  \sinh  \omega (t_f - t_i) } } \; e^{- \frac{m \omega}{2 \hbar} ( x^2 + y^2)  \coth  \omega (t_f - t_i)} \, 
e^{ \frac{ m \omega x y}{\hbar  \sinh  \omega (t_f - t_i) } } \;,
\eea
in agreement with the heat kernel for the harmonic oscillator given in eq. (27) of \cite{Khandekar}.

If we now also set
$\omega =0$,  we obtain the propagator for a free particle,
\bea
\label{heatfree}
K_E (y, t_f; x, t_i) = \sqrt{\frac{m }{ 2 \pi \hbar   (t_f - t_i) } } \; e^{- \frac{m }{2 \hbar} \frac{( x^2 + y^2) }{ (t_f - t_i)} } \, 
e^{ \frac{ m  x y}{\hbar   (t_f - t_i) } } = 
\sqrt{\frac{m }{ 2 \pi \hbar   (t_f - t_i) } } \; e^{- \frac{m }{2 \hbar} \frac{( x - y)^2 }{ (t_f - t_i)} } . \nonumber\\
\eea

\providecommand{\href}[2]{#2}\begingroup\raggedright\endgroup

\end{document}